\documentclass{amsart}

\usepackage{amsmath,amssymb}
\usepackage{graphicx}
\graphicspath{{figures/}}
\usepackage[hidelinks]{hyperref}

\title{Competitive and Complementary Tools}

\author{David C. Krakauer}
\address{Santa Fe Institute, 1399 Hyde Park Road, Santa Fe, NM 87501, USA}
\email{krakauer@santafe.edu}


\date{\today}

\begin{document}

\begin{abstract}
Humans have always externalized thought onto tools, from the tally and the
abacus to the map and, now, large language models. 
I model the agent, the tool, and the task as one dynamical system in which
competence (what the user retains) and reliance (what the user outsources) co-evolve, and find
that the outcome is bistable. Above a critical tool availability the competent state is
destroyed and competence collapses toward a low dependent floor as the user outsources completely. Lowering
availability does not reverse the collapse until a far lower threshold, so history of practice rather
than the current tool fixes the state. Two users with the same present access can therefore occupy opposite and lasting states, one competent and one dependent, decided only by which they built first. The collapse threshold depends jointly on the
competence a user brings to a task and on the tool's transparency, the fraction of its working a user can reconstruct. In the
case where an agent faces an uncertain goal, a tool can cause agency itself to transfer to the tool and the
human-agent becomes an agentic-instrument, irreversibly, because the tool's model is too large to
internalize. The model is tested against several independent data sets, including GPS and map use, arithmetic expertise, and language models. These results reframe how tools should be built, how
artificial intelligence is deployed, and what a tool-resistant education might require.
\end{abstract}

\maketitle

\section{Model and Theory}

\subsection{Complementary and competitive tools}

Two tools can serve the same task and leave their users in opposite states of competence. A map 
 leaves navigational knowledge that persists once the map is removed\cite{tolman,tversky}, whereas
satellite navigation, which computes routes and issues commands, is associated with
poorer formation of spatial memory \cite{javadi}. Early reports
describe analogous effects of language models on essay writing and on critical
thinking\cite{kosmyna,leeluther}.  Bergson made tool use central to his account of mind, suggesting that we are less Homo sapiens (wise humans) than Homo faber (tool using humans) and that intelligence is ``the faculty of manufacturing artificial objects, especially tools to make tools''\cite{bergson}. Archeologists have made the same point based on material culture, from Oakley's man the tool-maker to Leroi-Gourhan's treatment of technique as the medium of human evolution\cite{oakley,leroigourhan}. External memory is ancient technology, from clay
counting tokens to the knotted khipu\cite{schmandt,urton}, and the study of distributed and
extended cognition has shown that the unit performing a complex task is often a coupled
human--tool system \cite{hutchins,clark,zhang}. Here I investigate
what remains when a tool is taken away. I call a tool \emph{complementary} when
competence transfers to the user and persists after withdrawal, and \emph{competitive} when
the user grows more dependent the more it is used and tool-free competence is compromised. Capability is what tools add and
augmented intelligence is what complementary artifacts organize in the mind.

 A complementary tool does not merely get replicated inside the tool-user's mind but scaffolds an efficient internal representation. Complementarity is a means of enabling reverse-engineering of the logic of an artifact. An abacus expert calculates on a mental board that recruits
visuospatial cortex in place of verbal rehearsal\cite{hatano,stigler,frank,dehaene,hanakawa}, and
the chess master sees a few learned chunks where the novice searches the board square by
square\cite{chasesimon,gobetsimon,gobetclarkson,saariluoma}. This re-encoding, which allows the user to do more
with less, provides an operational definition of intelligence that goes beyond being more
capable. It is one turn of a recurrent feedback between the user's representation and
engineered artifact, a helix of exbodiment\cite{exbody,sawamura}, that iteratively improves both mind and artifact.

\subsection{Agent, tool, and task}

 The agent, the tool, and the environment are treated as a single class of mathematical object, an open transducer, each reading
inputs and writing outputs through an internal state. These are wired together so that the agent drives the
tool, the agent and tool both act on the environment, and the agent reads the result back
(shown schematically in Methods, Fig.~\ref{fig:loop}). A task is a goal to be reached in this
loop, and the transducer comprises three internal registers: memory $M$, operations $O$, and policy $\Pi$, the last of which also sets the goal $\Pi_{\rm goal}$ to pursue. The agent is the transducer that holds the policy register and sets the goal. Agent and tool
are therefore interchangeable by construction.

Two ensemble quantities summarize an agent. Competence $\iota\in[0,1]$ is the fidelity of the
agent's own model of the task, scored as the performance that remains when the tool is withdrawn but built and eroded under availability, through an empirical practice law\cite{newell,logan}, and
reliance $x\in[0,1]$ is the propensity to offload a control step. These two quantities track the two task-execution registers, memory $M$ and operations $O$, on which each step is either performed by the agent itself or offloaded to the tool. Competence is the pooled fidelity of $M$ and $O$ read out with the tool withdrawn, and reliance is the shared probability of offloading an $M$ or $O$ step. The split is one of engagement, whether the agent's own registers execute the step or the tool's registers do, and not one of tool presence, since the tool is available at level $a$ throughout. $\Pi$ is fixed by the human, the agent. Microscopically each step is a
Bernoulli choice, offloading with probability $x$, and competence changes by small increments that rise with the steps the agent performs itself and fall with the steps it offloads. Internal capacity is bounded and partitioned so that offloading frees a
 channel capacity\cite{baddeley,sweller}. Equations (\ref{eq:iota})--(\ref{eq:x}) are the
 mean-field limit of this stochastic loop as the number of control steps per unit
time grows (see Methods).  The parameter $\sigma$ is a logistic of slope $k$ and tool
availability $a\in[0,1]$ the
control parameter,
\begin{align}
\dot\iota &= r\,(1-x)(\iota_{\max}-\iota)\;-\;\mu\,x\,\iota, \label{eq:iota}\\
\dot x &= \beta\,a\,\sigma(\iota_c-\iota)\,(1-x)\;-\;\big(\alpha(1-a)+\epsilon\big)\,x.
\label{eq:x}
\end{align}
Competence increases with the fraction of steps the agent performs itself, $(1-x)$, converging on the ceiling $\iota_{\max}$ at
rate $r$, and is eroded by the fraction it offloads at rate $\mu$. Both terms act while the tool is available, so the split is one of engagement and not of tool presence. Reliance is favored when the tool is
available and competence has fallen below threshold $\iota_c$ (gain $\beta$). Reliance on the tool is
abandoned when the tool is unavailable and at an intrinsic base rate (rates $\alpha,\epsilon$).

The ceiling $\iota_{\max}$ is not a property of the user but of the tool. I call it the tool's \emph{transparency} $\tau\equiv\iota_{\max}\in[0,1]$, the fraction of the tool's function that the user can reverse-engineer internally from inspection. A transparent tool, such as a map or an abacus, exposes a mechanism that can be rebuilt in the mind, so $\tau$ is near one. An opaque tool, such as satellite navigation, conceals the computation that turns positions into routes, so there is little to internalize and $\tau$ is small. Transparency therefore bounds the competence that any amount of practice constructs in the mind, and  enters the dynamics through $\iota_{\max}$ in Eq.~(\ref{eq:iota}). All learning is tool mediated and a transparent tool reconstructs part of the tool's mechanism, up to $\iota_{\max}=\tau$. Offloaded capability occurs when the agent passes a problem to the tool without reconstructing it, counted by $x$, and experience atrophy $\mu x\iota$. Output that is reconstructed and verified is therefore an active cognitive process that enters through $(1-x)$.

\subsection{Bistability, hysteresis, and  collapse}

\begin{figure*}[t]
\centering
\includegraphics[width=\textwidth]{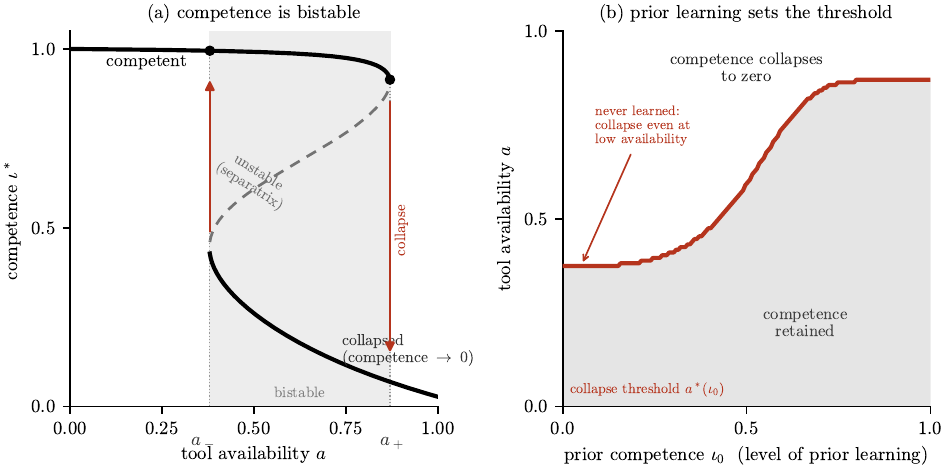}
\caption{\textbf{Competence collapses at a tipping point fixed by prior learning.}
\textbf{a}, Equilibrium competence $\iota^\ast$ as a function of tool availability $a$. A
competent branch and a dependent branch (competence driven toward zero) coexist between two
saddle-node folds $a_-$ and $a_+$ (shaded, bistable), separated by an unstable branch (the
separatrix). Raising availability past $a_+$ annihilates the competent state and competence
collapses (downward arrow), lowering availability restores it only below $a_-$ (upward arrow):
the loss is hysteretic. \textbf{b}, Running the full
(competence, reliance) dynamics from each prior competence $\iota_0$ at low initial
reliance partitions the plane of prior learning against tool availability $a$ into a region
where competence is retained (shaded) and a region where it collapses to the dependent state
(white). The collapse
threshold $a^\ast(\iota_0)$ (red) rises from $\approx0.38$ for an unlearned agent to
$\approx0.87$ for a well-trained one, below a level of prior learning, even a modestly
available tool drives competence to zero. Parameters in Methods.}
\label{fig:collapse}
\end{figure*}

These dynamics produce bistability. Reliance rises when the tool is available and competence 
slips, whereas competence is increased through work the user does on task and eroded by work  outsourced.
Because the temptation to offload increases
sharply once competence dips below a threshold, the coupled dynamics are nonlinear, and
over a range of tool availability $a$ they possess two stable states, a competent state and
a dependent state in which competence is driven toward zero\footnote{The dependent floor is
set by the intrinsic disengagement rate $\epsilon$, for the baseline parameters it is
$\iota^\ast\approx0.07$, with exact zero the idealized limit $\epsilon\to0$.} separated by an unstable
threshold, a separatrix (Fig.~\ref{fig:collapse}a). Raising availability past an upper fold
$a_+$ destroys the competent state, and competence collapses (Fig.~\ref{fig:collapse}a,
arrow). The collapse is hysteretic and lowering availability restores competence only below a
much lower fold $a_-$. Two users given the identical tool can therefore occupy opposite
states, set by their history of problem solving together with the transparency of the tool they are using.
The folds are not fixed values and  depend on the cost of acquiring and holding the
registers, and above a critical acquisition cost the competent state ceases to exist for any
availability or prior learning (Methods, Fig.~\ref{fig:cost}).

 Integrating the full system over levels of prior competence $\iota_0$ shows that the plane of prior learning against tool availability divides cleanly into a region
where competence is retained and a region where it collapses to the dependent state
(Fig.~\ref{fig:collapse}b). The boundary $a^\ast(\iota_0)$ is non-decreasing in prior
learning (flat at low $\iota_0$, then rising). An agent that never learned ($\iota_0$ near zero) collapses quickly, in the current model once availability
exceeds roughly $0.38$ of its maximum, whereas a well-trained agent persists until roughly $0.87$.
Prior learning therefore sets the level of tool availability a
person can withstand before the underlying competence is lost. Below a threshold of prior
learning, even an intermittently present tool carries competence to zero. The quantity one
might call ``analog education'' is what determines on which side of this critical value
a learner falls.

\begin{figure}[t]
\centering
\includegraphics[width=\linewidth]{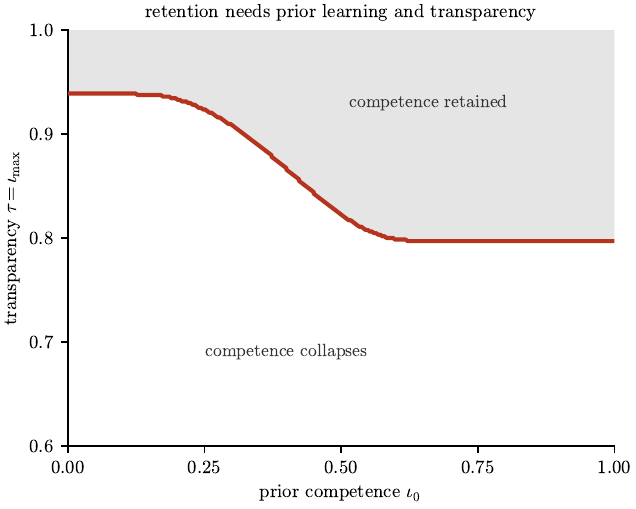}
\caption{\textbf{Retention requires both prior competence and transparency.}
Whether competence is retained, computed from the full dynamics at a fixed moderate
availability, over prior competence $\iota_0$ and the tool's transparency
$\tau=\iota_{\max}$ (the fraction of its function the user can reconstruct; the axis is
shown over the band where the transition occurs). The retained region (shaded) is
bounded on two sides: prior learning must be high enough that competence is built before
reliance takes over, and transparency must exceed a critical value, below which the
ceiling caps competence beneath the temptation threshold and no prior learning is
preserved. A transparent tool is retained from modest prior competence; as transparency
falls the required prior competence rises, and an opaque tool collapses for every
learner, however skilled. Figure~\ref{fig:collapse}b is the transparent slice
$\tau\!\approx\!1$ of this surface.}
\label{fig:transparency}
\end{figure}

The threshold is in fact a surface $a^\ast(\iota_0,\tau)$ in two variables, and Fig.~\ref{fig:collapse}b is its transparent slice $\tau\!\approx\!1$, where the tool's mechanism can be rebuilt and prior learning is decisive. As transparency falls the surface drops, because the ceiling $\iota_{\max}=\tau$ caps the competence that practice can restore once reliant on a tool. An agent with an opaque tool cannot rebuild from it (Fig.~\ref{fig:transparency}). This is the insights  from satellite navigation. An expert navigator who outsources route-finding to a device that conceals its computation will still lose spatial memory, because high prior competence and a near-zero ceiling leave nothing to re-internalize. A transparency below a critical value collapses competence for every level of prior learning, so the safe region of Fig.~\ref{fig:collapse}b shrinks from the top. Prior competence can only protect against tools transparent enough to be rebuilt. This is the difference between a map and a GPS.

\subsection{A typology of cognitive artifacts}

\begin{figure*}[t]
\centering
\includegraphics[width=\textwidth]{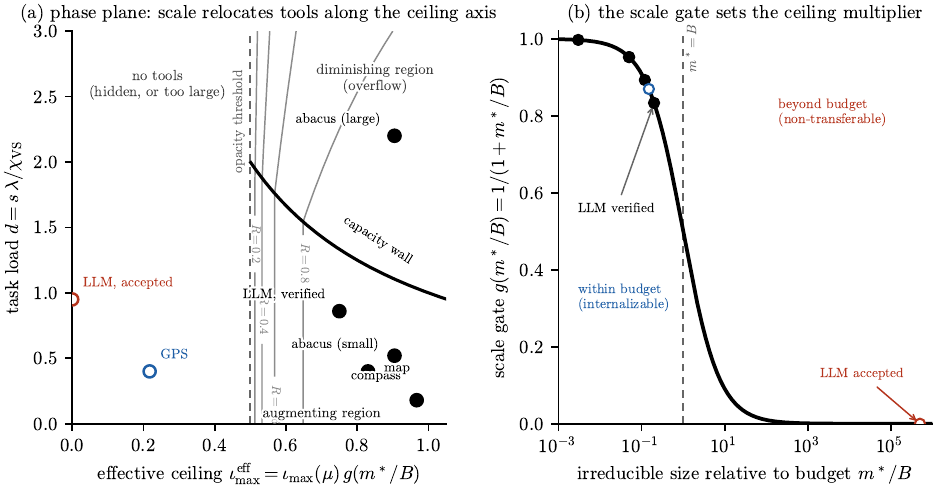}
\caption{\textbf{A map of competitive and complementary tools.}
\textbf{a}, Contours of retained gain $R$ after removal over the effective
interpretability ceiling $\iota_{\max}^{\rm eff}$ and the task's load $d$ on a
channel. An opacity threshold (vertical) and a capacity wall ($d=1/\iota_{\max}$) bound the
augmenting region; below the opacity threshold no tool transfers to the mind, and beyond
the capacity wall the internal representation overflows into the diminishing region.
Canonical tools are placed at their effective ceiling and coloured by
route: internalizable (black), concealed (blue), too large to internalize (red).
The language model appears twice, when reconstructed-and-verified it augments, whereas uncritical use
is non-transferable. \textbf{b}, The scale gate $g(m^\ast/B_{\rm a})=1/(1+m^\ast/B_{\rm a})$ that sets the ceiling
multiplier of internalization $m^\ast$ above which outsourcing is inevitable.}
\label{fig:phase}
\end{figure*}

The framework provides a natural means of classifying cognitive artifacts. In the long run
the gain that survives removal has a closed form (Methods) that places any tool on a plane
spanned by its transparency $\tau$, the ceiling on the surrogate a user can build from it,
and by how heavily the task loads the user's internal channels. The two key properties of
the model, prior competence and transparency, thus play distinct roles. Prior competence is
a property of the learner that decides which attractor a tool selects, whereas
transparency is a property of the tool that determines whether an attractor exists.
In lieu of data to calculate the placement of each artifact I estimate their positions within a calculable phase space
(Fig.~\ref{fig:phase}). Two boundaries organize the plane, with an opacity threshold (low $\tau$) below which
nothing transfers from tool to mind, and a capacity wall beyond which the internal representation overflows. The
compass enforces a single relation (the two points of a circle), the abacus, whose lattice-like board can be recoded into mental
imagery, and the map, whose metric compresses to topology, are all transparent and lie in the augmenting
region. Satellite navigation lies in the diminishing region because it conceals its
mechanism, which is to say its transparency $\tau$ is low. The language model, like a chess computer, can act either in the register of memory and operations, where it resembles a coach,
or in all three, where it becomes the player-agent and the human little more than its mechanical aid. Its effect on intelligence therefore splits with its mode of use. When its output is
reconstructed and verified, it transfers a method and augments intelligence. When its answers are accepted
uncritically, or when it describes a model far too large to internalize in the mind, it diminishes intelligence
(Fig.~\ref{fig:phase}, ``verified'' versus ``accepted''). Scale enters by lowering the
effective interpretability ceiling such that a sufficiently large model is non-transferable however
transparent its workings. 

\subsection{Population lock-in}

\begin{figure*}[t]
\centering
\includegraphics[width=\textwidth]{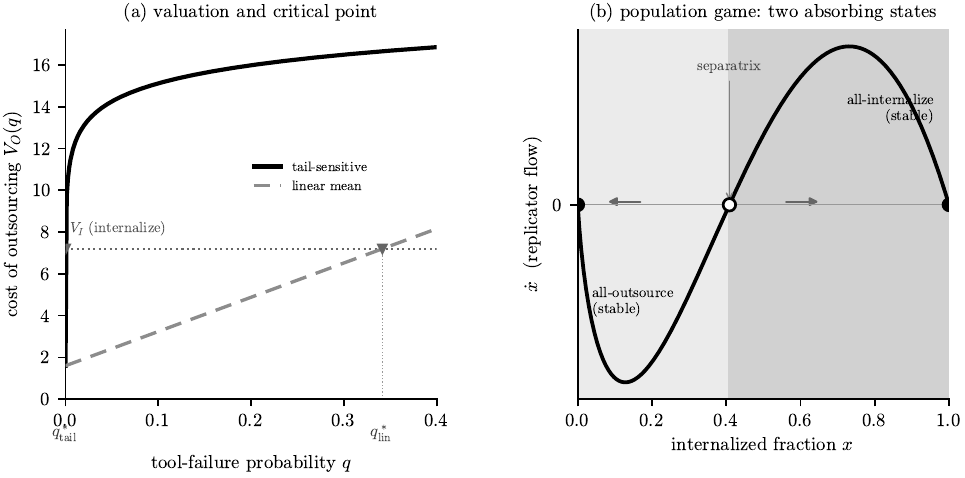}
\caption{\textbf{Collapse at the level of populations.}
\textbf{a}, The expected cost of outsourcing $V_O(q)$ under a tail-sensitive valuation
(black) crosses the internalizer's certain cost $V_I$ at a critical tool-failure probability
that the catastrophic tail pins near zero, whereas a risk-neutral mean (grey) lets it drift.
\textbf{b}, Replicator dynamics with availability coupled to the internalized fraction $y$:
two stable absorbing states (all-outsource, all-internalize) are separated by an unstable
separatrix.}
\label{fig:society}
\end{figure*}

The same instability scales to a population. A tool with a small user base becomes less reliable as users abandon the underlying
competence. With non-zero acquisition costs the population game has two absorbing states,
near-universal outsourcing and near-universal competence, separated by a tipping point
(Fig.~\ref{fig:society}). A society can be trapped in collective dependence even where broad
competence would be preferable.  The attractor it occupies is set by how much its members
learned before the tool became available. Individual hysteresis (critical periods) and collective lock-in are thus the same phenomenon at two scales governed by analogous dynamics.

\subsection{Agentic reversal}

I start with an assumed symmetry in which agent and tool are the same kind of transducer.
The agent holds the policy, but can lose it, allowing  for a condition involving the
exchange of roles.   An agent is by definition a transducer that
holds the policy $\Pi=(\Pi_{\rm path},\Pi_{\rm goal})$.  The challenge is that in complex settings, and hence when cognitive tools are sought, there is considerable goal uncertainty, the environmental goal state $G$ is unresolved, $H(G)>0$, and the human cannot
predict it. In terms of the transducer theory, the policy register is left open, and will be claimed by the transducer that can
resolve the goal. Writing $c_i^{M},c_i^{O},c_i^{\Pi}$ for the capacities of
transducer $i\in\{{\rm h},{\rm t}\}$ (human, tool) on memory, operations and rule-following,
if the goal is uncertain and the tool dominates,
\begin{equation}
H(G)>0,\qquad c_{\rm t}^{M}\ge c_{\rm h}^{M},\quad c_{\rm t}^{O}\ge c_{\rm h}^{O},\quad
c_{\rm t}^{\Pi}> c_{\rm h}^{\Pi},
\label{eq:reversal}
\end{equation}
then the hitting-time-minimizing holder of $\Pi$ is the tool (Methods). The tool becomes the
agent and the human, holding with discernible register advantage and unable to set goals, is reduced to
an instrument. Goal uncertainty in the presence of a powerful outsourcing option thus tends to displace humans from a policy-setting role towards becoming instruments of a computational agent.

What makes this more than a transient division of labour is that the role cannot be taken back.
Reclaiming agency would require internalizing the tool's goal-and-method model, which is typically one of irreducible
size $m^\ast_\Pi$. By the scale gate threshold the attainable ceiling is
$\iota_{\max}^{\rm eff}=\iota_{\max}\,g(m^\ast_\Pi/B_{\rm a})$, so for a policy far larger than the
acquisition budget, $m^\ast_\Pi\gg B_{\rm a}$, the gate closes, $\iota_{\max}^{\rm eff}\to0$, and the
retained gain is non-positive. Equivalently, the acquisition cost diverges and internalizing a representation
is economically dominated, $V_I\gg V_O$. The reversal is therefore an absorbing state of the
agency dynamics, and once the goal register has migrated to a tool whose model is too large to
rebuild, the change becomes difficult to reverse. This is a policy-register
counterpart of the competence collapse of Fig.~\ref{fig:collapse}. There the memory and
operations registers empty toward zero, whereas here, under a persistently uncertain goal and a
non-internalizable tool, the policy register is abandoned, and agency itself is
ceded to the ``cheaper'' choice.

\section{Empirical Predictions and Evidence}

\begin{table*}[t]
\centering\small
\resizebox{\textwidth}{!}{%
\begin{tabular}{@{}p{3.4cm}p{5.2cm}p{5.8cm}p{2.6cm}@{}}
\hline
model object & empirical claim & source & support\\\hline
$r$, $\iota_{\max}$ (Eq.~\ref{eq:iota}) & competence built by exercised fraction; ceiling bounded, training-dependent & manual skill tracks recent practice not exposure\cite{haslbeck}; bounded abacus transfer\cite{barner,barnerclass,du}; \cite{stigler,hatano} & strong; RCT \\
$\mu$ (Eq.~\ref{eq:iota}) & unexercised competence decays, faster for cognitive skills & skill-decay meta-analysis\cite{arthur}; disuse theory\cite{bjork} & strong; meta-analytic (Fig.~\ref{fig:atrophy}) \\
$\beta a\,\sigma(\iota_c-\iota)$ (Eq.~\ref{eq:x}) & offloading rises with availability and with low competence & expected-access memory effect\cite{sparrow}; offloading\cite{risko} & strong; experimental \\
positive feedback & reliance breeds reliance & offloading carry-over\cite{storm} & strong; experimental \\
bistability, folds $a_\pm$ & two stable states over an availability window & aviation dependency cycle\cite{casner,parasuraman} & moderate; indirect \\
divergence (Fig.~\ref{fig:divergence}) & more capable with tool, less without, in one agent & cognitive debt\cite{kosmyna}; \cite{gerlich}; session logs\cite{anthropic} & moderate; short-horizon \\
threshold $a^\ast(\iota_0)$ (Fig.~\ref{fig:collapse}b) & prior learning sets the tolerable availability & pilots\cite{haslbeck}; GPS longitudinal\cite{dahmani}; abacus experts\cite{stigler} & strong; longitudinal + cross-sectional \\
transparency $\tau=\iota_{\max}$ (Fig.~\ref{fig:transparency}) & opaque tool leaves nothing to rebuild, so competence is lost regardless of prior skill & GPS spatial-memory decline\cite{dahmani,javadi}; map vs GPS\cite{ishikawa} & strong \\
capacity wall & large task on partial surrogate overflows & bounded abacus transfer\cite{du,barnerclass} & moderate \\
population game (Fig.~\ref{fig:society}) & falling tool cost tips a population to dependence & task displacement in labor\cite{acemoglu} & strong; macro-empirical \\\hline
\end{tabular}}
\caption{Tool-based evidence. Each row lists a model object, the empirical requirement it implies, the studies that bear on it, and the strength of support. Three rows (bistability, divergence, and the capacity wall) have only correlational support and so double as predictions: a within-subject manipulation of availability, raised and then lowered, that should exhibit hysteresis, and a longitudinal trajectory in which capability and competence separate as reliance grows.}
\label{tab:evidence}
\end{table*}

The core assumptions of the model and its main results can now be checked against measurement.
Table~\ref{tab:evidence} organizes the published evidence. The first column describes a parameter or term of the dynamics
(Eqs.~\ref{eq:iota}--\ref{eq:x}) or a result the dynamics produce, such as the
bistability, the divergence, or the collapse threshold. The second column describes the meaning of the model terms.  The third column lists  independent studies drawn from skill retention, navigation, abacus expertise, cognitive
offloading, aviation, and labor economics.  The fourth column `grades' the support,
separating direct experimental or meta-analytic evidence from correlational or
short-horizon evidence, and flags any rows where the model predicts more than the
data currently can support. The subsections that follow consider the evidence for each of the model terms in more detail.

\subsection{Atrophy and the decay competence}

The atrophy rate $\mu$, the
erosion of competence by offloaded work, is the best measured of the parameters. A
meta-analysis of skill decay with disuse, pooling $189$ data points from $53$
studies, finds that loss grows from an effect size near zero immediately after
training to about $d=-1.4$ after a year of nonuse, and that cognitive, accuracy-based
skills decay faster than physical, speed-based ones\cite{arthur}. Fitting the
saturating form $-A(1-e^{-mt})$ to that trend recovers a cognitive decay roughly
twice the physical (half-lives near $48$ and $106$ days; Fig.~\ref{fig:atrophy}). The distinction between competence lost or
inaccessible, which hysteresis requires, is the storage-strength versus
retrieval-strength distinction of disuse theory\cite{bjork}. A competence whose
retrieval strength has decayed but whose storage strength survives is the latent
branch in the bifurcation analysis that re-establishes itself when the tool is withdrawn before atrophy has gone
too far.

\subsection{Acquisition ceilings}

The acquisition rate $r$ and its ceiling $\iota_{\max}$ are bounded by controlled
trials. Manual flying performance in airline pilots correlates with recent
hand-flying practice far more than with time since training, total experience, or
type rating\cite{haslbeck}. In the model competence is built by
the exercised fraction $(1-x)$ rather than by cumulative exposure. 

 The ceiling effect is supported by randomized trials
of mental abacus instruction, which show bounded transfer to untrained working
memory.  The largest gains appear after about three years and show no advantage
over alternative methods in a one-year classroom trial\cite{barner,barnerclass,du}.
The reliance dynamics, offloading that grows with availability $a$ and with a
temptation that rises once competence falls below $\iota_c$, are described in
cognitive-offloading literature. People who expect future access to information
retain less of it\cite{sparrow}, offloading is the use of external means to reduce
cognitive demand\cite{risko}, and reliance is self-reinforcing, since offloading on
one task raises it on the next\cite{storm}. This positive feedback, together with the sharp
response $\sigma$, produces two attractors.

\subsection{The collapse threshold}

\begin{figure}[t]
\centering
\includegraphics[width=\linewidth]{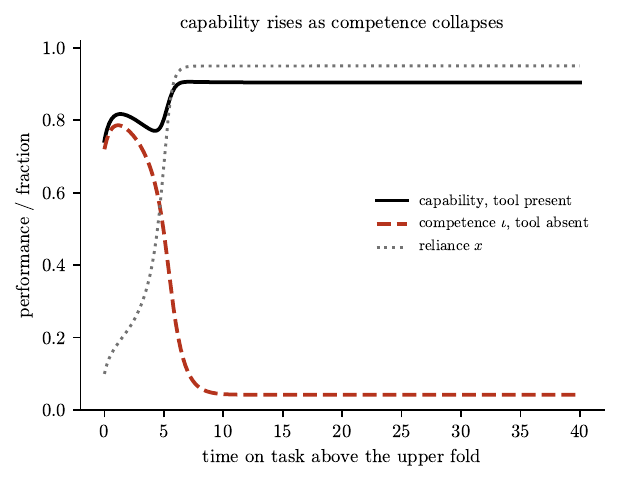}
\caption{\textbf{Capability and competence divergence.}
The full dynamics, Eqs.~(\ref{eq:iota})--(\ref{eq:x}), run at availability $a=0.95$
(above $a_+\approx0.87$). Reliance climbs, capability with the tool present stays high,
and competence (performance with the tool absent) collapses to the dependent floor.
The empirical anchors are the neural, behavioral, and session-log measurements of
refs.~\cite{kosmyna,gerlich,anthropic}. Computed from the model.}
\label{fig:divergence}
\end{figure}

\begin{figure}[t]
\centering
\includegraphics[width=\linewidth]{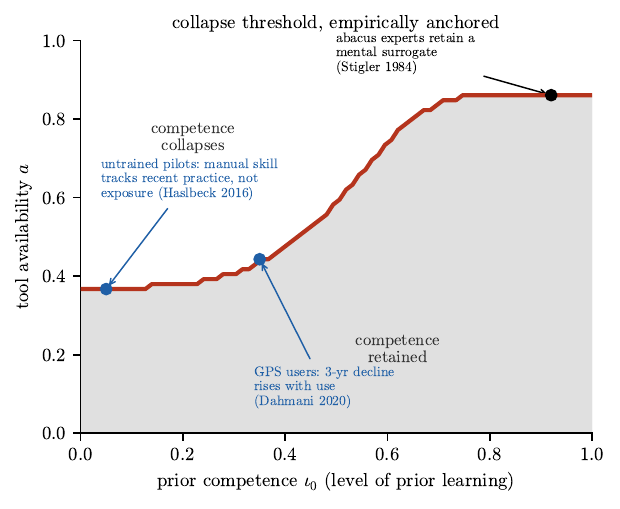}
\caption{\textbf{The collapse threshold.} The threshold
$a^\ast(\iota_0)$ (red), reproduced from the full dynamics (Methods), rises from
$\approx0.37$ for an unlearned agent to $\approx0.86$ for a well-trained one,
matching Fig.~\ref{fig:collapse}b. Anchors: the untrained low-availability collapse is
the automation-dependent pilot\cite{haslbeck} and the habitual satellite-navigation
user whose spatial memory declines with use\cite{dahmani}; the trained tolerance of
ubiquitous tools is the abacus expert who retains a mental surrogate\cite{stigler}.}
\label{fig:thresh}
\end{figure}

The central result, that capability and competence diverge in one agent above the
upper fold, has been observed neurally, behaviorally, and at the level of populations. Writers using a
language model show weaker neural connectivity and executive engagement than a
brain-only group\cite{kosmyna}, and self-reported critical thinking falls with
artificial-intelligence use, most in the youngest users\cite{gerlich}. An
analysis of roughly $400{,}000$ coding sessions separates a delegation pattern, fast
and low-error but with the least independent understanding, from a learning
pattern\cite{anthropic}. Run above the upper fold, the model reproduces this
divergence: reliance climbs, tool-present capability stays high, and tool-absent
competence collapses to the dependent floor (Fig.~\ref{fig:divergence}). The
hysteresis has been described as a vicious circle in aviation, where pilots with few
opportunities to hand-fly lose confidence and so prefer the automation, deepening the
loss\cite{casner,parasuraman}. This is the positive feedback drawn as a fold, and the
catastrophic-recovery accidents are its lower branch. The collapse
threshold's dependence on prior learning is anchored at both ends
(Fig.~\ref{fig:thresh}), since the untrained collapse at low availability describes pilots and
habitual satellite-navigation users, whose spatial memory declines over three
years in proportion to use, with the control that heavier users did not begin with a
poorer sense of direction, evidence that use drives the decline rather than the
reverse\cite{javadi,dahmani}, whereas highly trained experts tolerate near-ubiquitous tool use, including
abacus experts who retain a mental surrogate\cite{stigler,hatano}.

\subsection{Typology boundaries}

The two boundaries of the typology plane are supported by both the navigation and
abacus literatures. Satellite navigation sits in the diminishing region, its low effective ceiling fixed by the longitudinal decline of
hippocampal-dependent spatial memory with use\cite{dahmani,javadi}. The map sits
to its left because guided travelers acquire less route knowledge than map readers or
unaided navigators\cite{ishikawa}.  Map readers and satellite navigation users diverge in spatial memory at comparable exposure to the same routes,  which shows how the "use it or lose it" intuition is an oversimplification that can fail. This is because equal use should produce equal loss. The difference between these is explained in terms of transparency versus the frequency of use, and a large-task abacus exceeds the capacity
wall by task size becoming opaque \cite{du,barnerclass}
(Fig.~\ref{fig:phase}). Finally, the population result is the cognitive analog of the
task-based account of automation in labor economics, in which automation reallocates
tasks from labor to capital, always lowers the labor share, and can reduce labor
demand even while raising productivity when the displacement effect dominates
reinstatement\cite{acemoglu}. The displacement of a task from a worker is the
population analogue of the displacement of a register from an agent's competence to a
tool's, and the regime in which displacement dominates is the regime in which the
dependent attractor dominates.

\begin{figure*}[t]
\centering
\includegraphics[width=\textwidth]{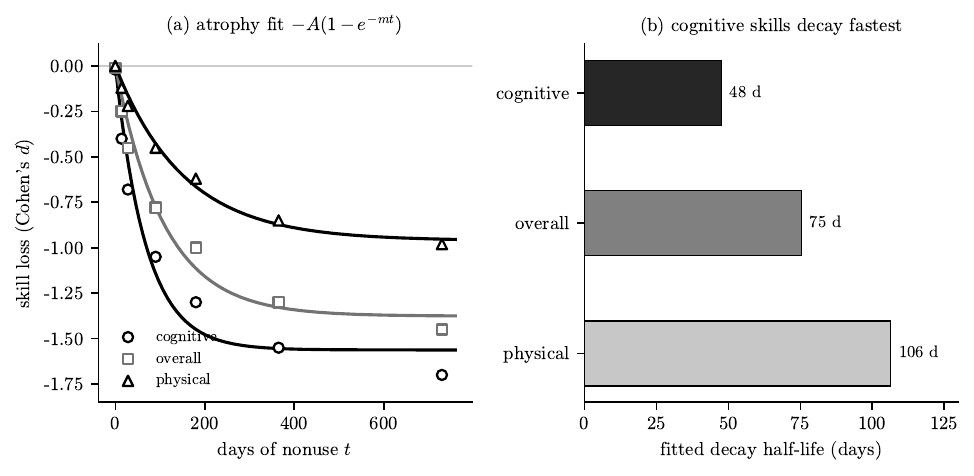}
\caption{\textbf{Atrophy rates and cognitive skill decay.}
\textbf{a}, Fit of the model's atrophy form $-A(1-e^{-mt})$ to the meta-analytic
skill-loss trend of ref.~\cite{arthur} as a function of days of nonuse, split by the
reported task-type moderator. Points are the published summary trend and curves are
least-squares fits. \textbf{b}, The fitted decay half-lives. Cognitive skills, the class a cognitive tool offloads, decay about twice as fast as physical skills, as captured by the term $\mu x\iota$.}
\label{fig:atrophy}
\end{figure*}

\section{Discussion}

The question of whether a tool strengthens or weakens its user is as old as technology itself, and it has been treated case by case, from the printed text to calculators and now AI. I have shown that both competence and reliance obey a law with a tipping-point, on one side of which a tool is complementary, and on the other it is competitive. The side is set by two features, how much the user learned before using the tool and how transparent the tool is, that is, how many of its features the user can reconstruct in their mind. A learner who built competence first can keep it, but only against a tool transparent enough to rebuild. Whereas a sufficiently opaque tool, such as satellite navigation, erodes competence however skilled the user. These formal results convert a diffuse cultural worry about growing dependence on technology into a number of quantifiable predictions. These include characterizing the path-dependent history of use through two thresholds and a critical period.  The model is able to classify navigation, aviation, arithmetic expertise, and the use of language models under one shared cognitive mechanism. The stakes are high because AI is now reaching billions of people, and it is an opaque tool appealing to users with limited prior competence,  thereby placing it in a regime that predicts competence collapse.

Someone who mastered arithmetic before the calculator was introduced possesses high prior competence above the collapse threshold, which in this paper, assuming a well trained agent,  sits at high availability ( $0.87$). Such a person keeps unaided arithmetic while offloading calculation, with retention best on low load operations that were practiced most. The model therefore predicts, rather than contradicts, the familiar observation that an adult trained in childhood can compute by hand decades later. The prediction that separates the model from a simple use it or lose it rule is revealed by the case involving a second person given the identical calculator at the identical availability without prior competence. The second learner starts below threshold (which for an untrained agent in the model is around $0.38$), and then converges on the dependent state. The same tool and the same access produce opposite and lasting (in)competence fixed by the order of learning. This is why arithmetic should be taught before the calculator and programming before using coding agents.  The model predicts the collapse of any cohort taught in the reverse order even when eventual tool use is identical in both.

The empirical record of tool use can be read as a partial measurement
of the parameters that distinguish complementary from competitive tools. The
atrophy rate $\mu$ is fixed by the skill-decay literature, which gives both its magnitude
and its dependence on task type, with cognitive registers decaying about twice as
fast as motor ones. The acquisition rate $r$ and the ceiling $\iota_{\max}$ are
bounded below by the recency dependence of manual skill and above by the limited
transfer of abacus training. The reliance gain $\beta a$ and the temptation threshold
$\iota_c$ appear in the offloading studies, in which expected access lowers
retention and offloading on one task raises it on the next. 

The variables of the model can be treated in the same way. Competence $\iota$ has been observed to fall as reliance $x$ rises within a single agent, and the collapse threshold
$a^\ast(\iota_0)$ shifts with prior learning across pilots, navigators, and abacus
experts. No previous study has measured these quantities jointly in one system,
and the model offers a way to connect literatures that otherwise appear unrelated but that bear on the same cognitive processes.

The model predicts that the
same task, tool, and population should exhibit two distinct steady states of competence
at one fixed availability. These are reachable by different histories, so that a within-subject
design that raises availability to destroy competence and then lowers it should trace a
hysteresis loop with two folds $a_-$ and $a_+$  separated by a gap. In the model there is an upper fold with availability $0.87$ where competence is lost down to a lower fold near $0.38$ where it returns. Regardless of the precise values of these thresholds, the intuition for the result is that recovery requires far more than simply abstaining from tool use. It
predicts that competence and capability separate at a rate set by $\mu$ and saturate at a floor fixed
by $\beta,\alpha,\epsilon$. It predicts a critical period, in that the order of two
identical curricula, one introducing the tool before competence is built and one after,
should select opposite long-run attractors. And it predicts a capacity wall, a task
size beyond which a partially internalized surrogate overflows and post-removal
performance drops below the unaided baseline, locating the wall at $d\approx
1/\iota_{\max}^{\rm eff}$.  Treating
artifacts as coupled dynamical systems rather than as static affordances opens an
empirical field whose task is to measure a tool, task, and learner
through the full loop of acquisition, reliance, and removal.

 Ubiquitous, frictionless artificial intelligence is a clear instance of the high-availability regime in which competence is
bistable. A generation that encounters these tools before acquiring competence begins below the
threshold of prior learning and is carried to the collapsed state. This is not through any failure
of the tool, which is working perfectly, but through the geometry of the dynamics.

The two results of the paper therefore map onto two important decisions that society and its educators must make. The first is that because prior competence on a tool sets which attractor a learner occupies, there is a decision to be made about sequencing.  Competence must be built above the separatrix first, with the tool introduced second as a complement, so that the learner can reconstruct and verify its output and land in the augmenting attractor. The reverse order lands users in the collapsed state, which is why the order of the curriculum, and not merely its content, matters. The second is that transparency decides whether an attractor exists at all, and is therefore a decision about which tools to use. A transparent tool, one whose working a learner can rebuild in the mind, such as a map or an abacus, can be allowed early because practice on it generates competence. Whereas an opaque tool, such as a device that conceals its computation or a model too large to internalize, removes the retained state for every learner however skilled. Such tools should be used long after competence is secure. An education that is robust to powerful tools is one that builds competence before reliance and allows, in the learning period, only tools students could in principle do without. The cultural evolution of humanity is inseparable from the use of tools but the choice of tool matters. This distinction should inform education in the age of AI as well as all those independent-minded thinkers compelled to ask whether we want humanity to preserve agency or to become the embodied tools of an agentic tool.

\pagebreak 

\appendix
\section{Methods}
\small

\subsection{Transducers and their mean field limits}

\begin{figure*}[t]
\centering
\includegraphics[width=\textwidth]{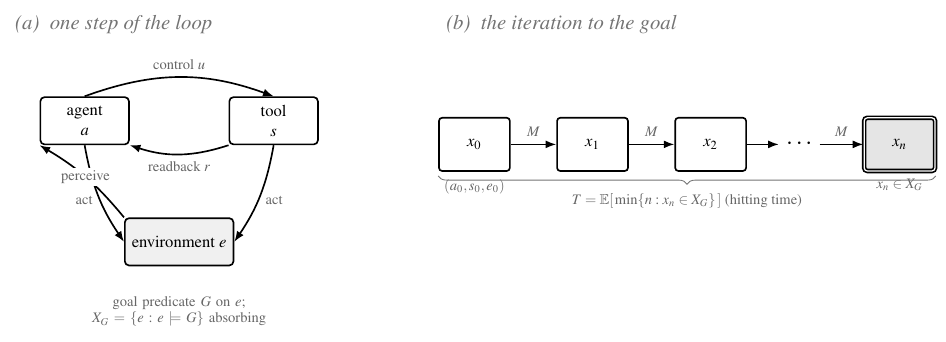}
\caption{\textbf{The task loop and agent--tool symmetry.}
\textbf{a}, Agent, tool and environment are open transducers wired into a loop: the agent
emits a control $u$ and reads back a result $r$, and agent and tool both act on the
environment, whose goal states $X_G$ are absorbing. Nothing distinguishes the agent except
possession of the policy register and the goal, so agent and tool can exchange roles.
\textbf{b}, Unrolled, the joint state $x_n=(a,s,e)$ advances under the composite kernel $M$
until it reaches $X_G$; performance is the hitting time $T$.}
\label{fig:loop}
\end{figure*}

The agent, tool and environment are modelled as open transducers, each a tuple
$(S,I,U,\delta,\lambda)$ of a state space $S$, inputs $I$, outputs $U$, a stochastic
transition kernel $\delta$ that advances the state under an external input, and an output map
$\lambda$. Wiring the three into a loop, the agent emits a control $u$ that drives the tool
and acts on the environment, the tool acts on the environment, and the agent reads the result
back (Fig.~\ref{fig:loop}a), makes the joint state $x=(a,s,e)$ a Markov chain on the product
space, advancing under a composite kernel $M$. A task is a goal $G$ on the
environment whose states $X_G$ are absorbing. Performance is the hitting time
$T=\mathbb{E}[\min\{n:x_n\in X_G\}]$, and
the fundamental matrix of the absorbing chain\cite{kemeny} (Fig.~\ref{fig:loop}b). The work a
task demands resolves onto three registers: memory $M$ (state held), operations $O$
(transformations applied) and policy $\Pi$ (which transformation, toward which goal).
 The loop is  symmetric in agent and tool, and their roles can exchange by design.
 
A control step employs the two task-execution registers, memory $M$ and operations $O$. The
policy register $\Pi$ selects which step to take and toward which goal. The goal is held by the agent
so it does not appear as a state variable in Eqs.~(\ref{eq:iota})--(\ref{eq:x}). In principles
 $M$ and $O$ are separate competences $\iota_M,\iota_O$ with their own offload
probabilities but because both obey the same rules and are driven by the same tool
availability $a$, the symmetric case collapses to a single $\iota=w\iota_M+(1-w)\iota_O$ and a
single $x$ (any fixed task-dependent weight $w$ rescales $r,\mu$ without changing the bifurcation
structure). Competence is the pooled fidelity of
$M$ and $O$. Reliance is the pooled offload probability across $M$- and $O$-steps, and
in the agency analysis,  the policy register
 can transfer to the tool. The ceiling $\iota_{\max}$ is the fidelity of the tool-mediated mental representation the agent builds by constructed through tool use, and it bounds the growth term. The model describes agents who always operate with tools and calculates what is retained. I do not consider agents who never use tools.
 
Competence and reliance are treated as the  limit of a continuous-time Markov jump process.
 In an interval $dt$ the agent attempts $n\,dt$ steps and each is offloaded with
probability $x$ otherwise performed by the agent itself. The internal capacity is bounded and partitioned so
that offloading frees a channel\cite{baddeley,sweller}. Use of a complementary tool raises
competence by a constant proportional to $(\iota_{\max}-\iota)$ and an offloaded step lowers it by a
constant proportional to $\iota$ and the offload probability jumps up when the tool is available
and competence has fallen below $\iota_c$, down otherwise, all increments of size $O(1/n)$. As
$n\to\infty$ (many small control steps per unit time) the rescaled process converges, by the
law of large numbers for density-dependent Markov jump processes\cite{kurtz}, to the
deterministic solution of Eqs.~(\ref{eq:iota})--(\ref{eq:x}), with Gaussian fluctuations of
order $n^{-1/2}$ about it. Equations~(\ref{eq:iota})--(\ref{eq:x}) are  the 
$n\to\infty$ mean field of the stochastic loop. The same
construction with rates depending on the population density, rather than state,
yields the replicator dynamics used for the population game. The convergence and the
$n^{-1/2}$ fluctuation bound hold on finite time horizons, so at finite $n$ the competent state is
metastable rather than absolutely stable: noise can carry the system across the separatrix, and the
sharp hysteresis of Eqs.~(\ref{eq:iota})--(\ref{eq:x}) is the $n\to\infty$ idealization of a
crossing whose expected time is long but finite.

The quasi-steady reliance $x^\ast(\iota,a)=D/(D+B)$ with $D=\beta a\,\sigma(\iota_c-\iota)$ and
$B=\alpha(1-a)+\epsilon$ is exactly the $x$-nullcline of Eqs.~(\ref{eq:iota})--(\ref{eq:x}), so
every equilibrium of the full two-dimensional system is a zero of the  field
$F(\iota;a)=r(1-x^\ast)(\iota_{\max}-\iota)-\mu x^\ast\iota$. Stability is taken from the full $2\times2$ Jacobian rather than from
$\partial F/\partial\iota$: at the baseline (Table~\ref{tab:par}) with $a=0.6$ the three
equilibria $\iota^\ast=0.99,\,0.67,\,0.19$ have eigenvalues $(-1.12,-0.39)$, $(-2.22,+0.45)$
and $(-1.06,-2.35)$, marking the competent and dependent branches stable and the middle branch
a saddle (the separatrix). Continuing the
equilibria in $a$ (bracketing sign changes of $F$ on a fine $\iota$-grid and refining)
returns the competent branch (highest stable root), the dependent branch (lowest stable
root) and the separatrix (unstable root). Where the equilibrium count changes are the
saddle-node folds $a_-$ and $a_+$. For the baseline parameters (Table~\ref{tab:par})
$a_-\approx0.38$ and $a_+\approx0.87$, and the dependent branch has competence
$\iota^\ast\approx0.07$ (Fig.~\ref{fig:collapse}a).

\subsection{Agentic Reversals}

\begin{figure*}[t]
\centering
\includegraphics[width=\textwidth]{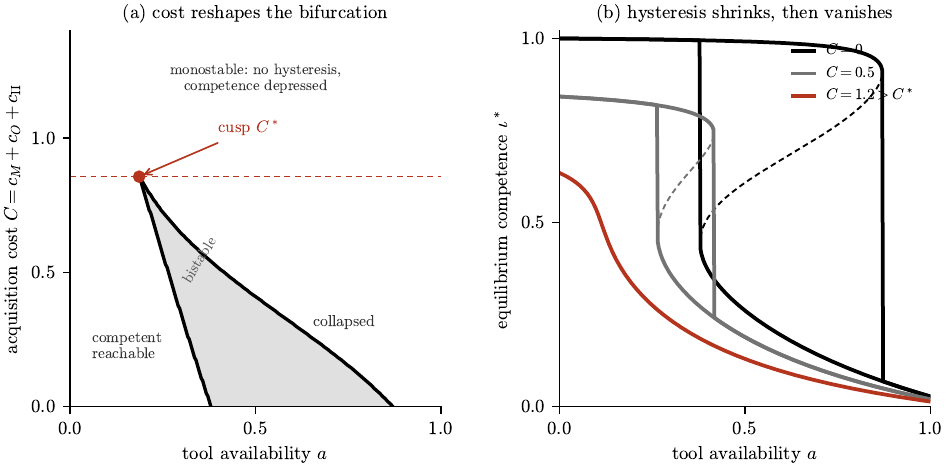}
\caption{\textbf{Acquisition cost reshapes the competence bifurcation.}
\textbf{a}, Two-parameter bifurcation in tool availability $a$ and composite acquisition cost
$C=c_M+c_O+c_\Pi$. The saddle-node fold curves bound a bistable tongue (shaded) that closes at
a cusp $C^\ast$. Below $C^\ast$ the collapse threshold slides to lower availability as cost
rises. Above $C^\ast$ the dynamics are monostable, with no hysteresis and depressed
competence. \textbf{b}, Equilibrium competence $\iota^\ast(a)$ at three costs: the hysteresis
loop (stable branches solid, separatrix dashed) shrinks and shifts left with cost, and for
$C>C^\ast$ becomes a single branch that falls continuously and lies below the competent state
at every availability.}
\label{fig:cost}
\end{figure*}

Let $C=c_M+c_O+c_\Pi\ge0$ be the composite cost of acquiring the memory, operations
and policy registers, and let cost enter the slow field through two channels, a throttle on
acquisition (costly registers are built more slowly) and a drain on what is held (costly
registers must be maintained, an upkeep paid even with the tool absent),
\begin{equation}
\dot\iota=\frac{r}{1+\kappa C}\,(1-x)(\iota_{\max}-\iota)\;-\;\mu\,x\,\iota\;-\;\nu\,C\,\iota,
\label{eq:cost}
\end{equation}
with $\kappa,\nu\ge0$; $C=0$ recovers Eqs.~(\ref{eq:iota})--(\ref{eq:x}) exactly. Repeating
the reduction and continuation with $F(\iota;a,C)$ turns the single fold diagram into a
two-parameter bifurcation in $(a,C)$ (Fig.~\ref{fig:cost}). For $0\le C<C^\ast$ the system is
still bistable, but the window slides toward lower availability as cost rises---acquisition
cost lowers the collapse threshold $a^\ast$, the mirror image of prior learning, which raises
it. At a critical cost $C^\ast$ (for $\kappa=0.5,\nu=0.3$, $C^\ast\approx0.86$) the two
folds meet at a cusp and above it the competent and dependent branches are no longer
simultaneously stable, hysteresis is gone, and equilibrium competence both declines
continuously with availability and is depressed at every availability, so that no level of
prior learning recovers it. The register decomposition makes the link to the rest of the
paper where the policy term $c_\Pi$ is the acquisition cost of the tool's goal-and-method
model, the same quantity the scale gate measures ($c_\Pi\!\sim\!m^\ast_\Pi/B_{\rm a}$), so a tool whose
policy cannot be internalized drives $C$ past $C^\ast$.

For Fig.~\ref{fig:collapse}b the full system (\ref{eq:iota})--(\ref{eq:x}) is integrated to
$t=260$ from each prior competence
$\iota_0$ on a $160$-point grid and each availability $a$ on a $140$-point grid, with low
initial reliance $x_0=0.05$ representing an agent newly meeting the tool. The recorded final
competence partitions the $(\iota_0,a)$ plane, the collapse threshold $a^\ast(\iota_0)$ is the
largest $a$ for which final competence exceeds one half. It rises monotonically from
$\approx0.38$ at $\iota_0\!\to\!0$ to $\approx0.87$ at high $\iota_0$, the central result.
This panel fixes the ceiling at $\iota_{\max}=\tau=1$ (a fully transparent tool), so the only
variable is prior learning. Figure~\ref{fig:transparency} instead varies the transparency
$\tau=\iota_{\max}$ at fixed moderate availability and records whether the trajectory settles on
the competent branch (final competence above $\iota_c$); the initial competence is capped at the
ceiling, $\iota(0)=\min\{\iota_0,\tau\}$, since a user cannot start above what the tool exposes.
Because the ceiling caps competence below the temptation threshold once $\tau<\iota_c$, a
transparency below a critical value collapses competence for every $\iota_0$, the formal version
of the satellite-navigation case.

\subsection{Typology of tools}
Figure~\ref{fig:phase} is the long-training limit. After removal the agent runs its internal representation
at fidelity $\iota_{\max}$. Below a residual
floor $C_\infty$ nothing transfers, so the transferable fraction is
$\phi(\iota_{\max})=\max\{0,(\iota_{\max}-C_\infty)/(1-C_\infty)\}$, and when the model's
load exceeds capacity it is coarsened by $\min\{1,1/(\iota_{\max}d)\}$, where
$d=s\,\ell/\chi$ is task load ($s$ subtasks at per-subtask load $\ell$) relative to channel
capacity $\chi$. The effective transferable
competence is $\xi=\phi(\iota_{\max})\min\{1,1/(\iota_{\max}d)\}$, and the retained gain is
$R=(T_{\rm bare}-T_3)/(T_{\rm bare}-T_{\rm tool})$ with task reward scale $W$, bare and tooled
times $T_{\rm bare}=W/p_{\rm bare}$, $T_{\rm tool}=W/p_{\rm tool}$, and
$T_3=W/(p_{\rm bare}+\xi(p_{\rm tool}-p_{\rm bare}))$. Scale enters through the effective
ceiling $\iota_{\max}^{\rm eff}=\iota_{\max}\,g(m^\ast/B_{\rm a})$, with the gate
$g(m^\ast/B_{\rm a})=1/(1+m^\ast/B_{\rm a})$ in the irreducible internalized size $m^\ast$ relative to the
acquisition budget $B_{\rm a}$, opacity is reached either by concealment (low intrinsic ceiling) or
by scale ($m^\ast\!\gg\!B_{\rm a}$, $g\to0$), which is why a large language model is
non-transferable. Parameter values and the per-tool $(\iota_{\max}^{\rm eff},d)$ placements are
in Table~\ref{tab:toolmap}.

\subsection{Population dynamics.}
For Fig.~\ref{fig:society} an agent chooses between outsourcing and internalizing, and the
internalized fraction $y$ of the population (distinct from the single-agent reliance $x$)
obeys replicator dynamics driven by the difference between the two strategies' costs:
\begin{align}
\dot y &= y\,(1-y)\,\big(V_O(q(y)) - V_I\big), \label{eq:replicator}\\[2pt]
V_O(q) &= \frac{1}{\rho}\,\log\!\Big[(1-q)\,e^{\rho(T_{\rm tool}+C_{\rm op})}
            + q\,e^{\rho\Lambda}\Big], \label{eq:VO}\\[2pt]
V_I &= T_{\rm tool} + C_\infty + (1-\zeta)K, \qquad
q(y) = q_0 + (q_{\max}-q_0)\,y. \label{eq:VI}
\end{align}
Here $V_O$ is the certainty-equivalent cost of the outsourcing lottery under risk aversion
$\rho$, with tool-failure probability $q$, operation cost $C_{\rm op}$, and catastrophic
failure cost $\Lambda$; $V_I$ is the certain cost of internalizing, with acquisition cost $K$
and retained fraction $\zeta$. The failure rate rises with the internalized
fraction ($q_{\max}>q_0$), and since costs are penalties the internalizers grow when
$V_O>V_I$. For acquisition cost in the trapping range, Eq.~(\ref{eq:replicator}) has two stable
absorbing states, $y=0$ (near-universal outsourcing) and $y=1$ (near-universal competence),
separated by an unstable interior fixed point---the population-level image of the individual
fold. Parameter values are in Table~\ref{tab:popgame}, the residual floor $C_\infty$ and tooled
time $T_{\rm tool}$ are illustrative economic scales set independently of their tool-map
values.

\subsection{From human-agents to agentic-isntruments}
Agency in the transducer loop is carried by the policy register and the agent is the transducer that holds
$\Pi=(\Pi_{\rm path},\Pi_{\rm goal})$, with $\Pi_{\rm goal}$ the selection of the goal $G$. When $G$ is supplied, and the assignment is fixed,
goal uncertainty becomes an issue, $H(G)>0$, with $H$ the entropy of the agent's posterior over candidate
goals. Let $c_i^{M},c_i^{O},c_i^{\Pi}\ge0$ be the capacities of transducer
$i\in\{{\rm h},{\rm t}\}$ on memory, operations and policy, where $c_i^{\Pi}$ measures both
goal resolution (the rate at which $i$ reduces $H(G)$) and rule-following (selection of
transformations).  The hitting time is non-increasing in each capacity with
$T=(I-Q)^{-1}\mathbf{1}$ for the within-task kernel $Q$, raising a capacity sends
$Q\mapsto Q'$ with $Q'\le Q$ entrywise (probability mass shifted toward the absorbing goal),
and since $(I-Q)^{-1}=\sum_{k\ge0}Q^{k}$ is entrywise monotone in $Q$, $T'\le T$ componentwise.

The policy register is allocated to the hitting-time-minimizing transducer
under goal uncertainty, $\Pi^\ast=\arg\min_{i}\mathbb{E}[T\mid\Pi\ \text{at}\ i]$. 
Whenever condition~(\ref{eq:reversal}) holds, $H(G)>0$ and $c_{\rm t}^{M}\ge c_{\rm h}^{M}$,
$c_{\rm t}^{O}\ge c_{\rm h}^{O}$, $c_{\rm t}^{\Pi}>c_{\rm h}^{\Pi}$, we find that
$\Pi^\ast={\rm t}$, and the tool becomes the agent and the human a subordinate transducer.
Reclaiming agency requires a mental representatin of the tool's policy model, of irreducible size
$m^\ast_\Pi$, by the scale gate the attainable fidelity ceiling is
$\iota_{\max}\,g(m^\ast_\Pi/B_{\rm a})$, so for $m^\ast_\Pi\gg B_{\rm a}$ the ceiling and the retained gain
vanish, and the acquisition cost $K(m^\ast_\Pi)$ diverges so that $V_I\gg V_O$. The allocation
$\Pi^\ast={\rm t}$ is a fixed point that cannot be exited once $m^\ast_\Pi>B_{\rm a}$. Hence under a
persistently uncertain goal and a non-internalizable tool this is the policy-register analogue of
the competence collapse (Fig.~\ref{fig:collapse}), in which the memory and operations registers
empty.

\paragraph{Parameters, code and data.}
\begin{table}[h]\centering\small
\begin{tabular}{@{}lll@{}}
\hline
symbol & meaning & value\\\hline
$r$ & acquisition rate & $1.0$\\
$\mu$ & atrophy rate & $1.2$\\
$\beta$ & adoption gain with availability & $3.0$\\
$\alpha$ & abandonment gain with unavailability & $1.0$\\
$\epsilon$ & intrinsic disengagement & $0.10$\\
$\iota_{\max}$ & competence ceiling & $1.0$\\
$\iota_c$ & temptation threshold & $0.50$\\
$k$ & logistic slope & $12.0$\\
$\kappa$ & acquisition-cost throttle (Eq.~\ref{eq:cost}) & $0.5$\\
$\nu$ & maintenance-cost drain (Eq.~\ref{eq:cost}) & $0.3$\\\hline
\end{tabular}
\caption{Baseline parameters for Eqs.~(\ref{eq:iota})--(\ref{eq:x}), $\kappa,\nu$ govern the cost extension Eq.~(\ref{eq:cost}), with baseline $C=0$.}
\label{tab:par}
\end{table}

\begin{table*}[t]\centering\small
\begin{tabular}{@{}lll@{}}
\hline
symbol & meaning & value\\\hline
$W$ & task reward scale & $20$\\
$p_{\rm bare},\,p_{\rm tool}$ & success rate bare / tooled & $0.10,\,0.95$\\
$C_\infty$ & residual floor (opacity threshold) & $0.5$\\
$B_{\rm a}$ & acquisition budget (blocks) & $10^{3}$\\
\hline
\multicolumn{3}{@{}l}{\emph{per-tool} $(\iota_{\max}^{\rm intr},\,m^\ast,\,d)$:}\\
compass & $(0.97,\ 3,\ 0.18)$ & internalizable\\
abacus (small / large task) & $(0.95,\ 50,\ 0.52/2.20)$ & overflow at large $d$\\
map & $(0.93,\ 120,\ 0.40)$ & internalizable\\
LLM, verified & $(0.90,\ 200,\ 0.86)$ & $m^\ast<B_{\rm a}$, $g\!\approx\!1$\\
LLM, accepted & $(0.90,\ 5\!\times\!10^{8},\ 0.95)$ & $m^\ast\!\gg\!B_{\rm a}$, $g\!\to\!0$\\
GPS & $(0.25,\ 150,\ 0.40)$ & concealment\\\hline
\end{tabular}
\caption{Tool-map parameters (Fig.~\ref{fig:phase}), $\iota_{\max}^{\rm eff}=\iota_{\max}^{\rm intr}\,g(m^\ast/B_{\rm a})$ places each tool, and $T_{\rm bare}=W/p_{\rm bare}=200$, $T_{\rm tool}=W/p_{\rm tool}\approx21$.}
\label{tab:toolmap}
\end{table*}

\begin{table}[h]\centering\small
\begin{tabular}{@{}lll@{}}
\hline
symbol & meaning & value\\\hline
$T_{\rm tool},\,C_{\rm op}$ & tooled time, operation cost & $1.0,\,0.6$\\
$C_\infty,\,\zeta$ & residual floor, retained fraction & $0.2,\,0$\\
$\rho,\,\Lambda,\,K$ (panel a) & risk aversion, failure cost, acquisition & $0.8,\,18,\,6$\\
$\rho,\,\Lambda,\,K$ (panel b) & "" & $0.5,\,14,\,10$\\
$q_0,\,q_{\max}$ & failure rate at $y=0,1$ & $0,\,0.6$\\\hline
\end{tabular}
\caption{Population-game parameters (Fig.~\ref{fig:society}), these economic scales are set independently of the tool-map values of $C_\infty,T_{\rm tool}$.}
\label{tab:popgame}
\end{table}
All figures generated from the equations and parameters in Methods. Code reproducing every figure, are available as a
Wolfram-notebook. 
\normalsize
\normalsize

\section*{Funding}
This work was supported by the Templeton World Charity Foundation, Inc. (funder DOI 501100011730) under grant \url{https://doi.org/10.54224/20650}, and by the Robert Wood Johnson Foundation (grant no.\ 81366), ``Using Emergent Engineering for integrating complex systems to achieve an equitable society.''

\section*{Data accessibility}
No new data were generated. The empirical figures fit model functional forms to, or place model coordinates against, published summary statistics from the cited studies. The source values used in Fig.~\ref{fig:atrophy} are the meta-analytic trend of ref.~\cite{arthur}, and the placements are listed in Table~\ref{tab:evidence}.

\section*{Competing interests}
The author declares no competing interests.

\end{document}